\documentclass[12pt,preprint]{aastex}

\renewcommand {\deg}   {\mbox{$^\circ$}}

\newcommand   {\kms}   {\mbox{km\,s$^{-1}$}}
\renewcommand {\ga}    {\mbox{\rlap{\hbox{\lower5pt\hbox{$\sim$}}}\hbox{$>$}}}
\renewcommand {\la}    {\mbox{\rlap{\hbox{\lower5pt\hbox{$\sim$}}}\hbox{$<$}}}

\begin{document}



\def\kms {\hbox{km{\hskip0.1em}s$^{-1}$}} 
\def\msol{\hbox{$\hbox{M}_\odot$}}
\def\lsol{\hbox{$\hbox{L}_\odot$}}
\def\kms{km s$^{-1}$}
\def\Blos{B$_{\rm los}$}
\def\etal   {{\it et al. }}                     
\def\psec           {$.\negthinspace^{s}$}
\def\pasec          {$.\negthinspace^{\prime\prime}$}
\def\pdeg           {$.\kern-.25em ^{^\circ}$}
\def\degree{\ifmmode{^\circ} \else{$^\circ$}\fi}
\def\ee #1 {\times 10^{#1}}          
\def\ut #1 #2 { \, \textrm{#1}^{#2}} 
\def\u #1 { \, \textrm{#1}}          
\def\nH {n_\mathrm{H}}

\def\ddeg   {\hbox{$.\!\!^\circ$}}              
\def\deg    {$^{\circ}$}                        
\def\le     {$\leq$}                            
\def\sec    {$^{\rm s}$}                        
\def\msol   {\hbox{$M_\odot$}}                  
\def\i      {\hbox{\it I}}                      
\def\v      {\hbox{\it V}}                      
\def\dasec  {\hbox{$.\!\!^{\prime\prime}$}}     
\def\asec   {$^{\prime\prime}$}                 
\def\dasec  {\hbox{$.\!\!^{\prime\prime}$}}     
\def\dsec   {\hbox{$.\!\!^{\rm s}$}}            
\def\min    {$^{\rm m}$}                        
\def\hour   {$^{\rm h}$}                        
\def\amin   {$^{\prime}$}                       
\def\lsol{\, \hbox{$\hbox{L}_\odot$}}
\def\sec    {$^{\rm s}$}                        
\def\etal   {{\it et al. }}                     

\def\xbar   {\hbox{$\overline{\rm x}$}}         

\slugcomment{Submitted to ApJL}
\shorttitle{}
\shortauthors{}

\title{Massive Star Formation in the Molecular
Ring Orbiting the Black Hole at the Galactic Center}
\author{F. Yusef-Zadeh\altaffilmark{1},
J. Braatz\altaffilmark{2},
M. Wardle\altaffilmark{3},
D. Roberts\altaffilmark{1}}

\altaffiltext{1}{Department of Physics and Astronomy,
Northwestern University, Evanston, Il. 60208
(zadeh@northwestern.edu)}
\altaffiltext{2}{National Radio Astronomy Observatory, 520 Edgemont Road, 
Charlottesville, VA 22903
(jbraatz@nrao.edu)}
\altaffiltext{3}{Department of Physics, Macquarie University, Sydney NSW 2109,
Australia (wardle@physics.mq.edu.au)}

\begin{abstract} 

A ring of dense molecular gas extending 2-7 pc orbits the supermassive 
black hole Sgr A* at the center of our Galaxy.  Using the Green Bank 
Telescope, we detected water maser lines and both narrow (0.35 km 
s$^{-1}$) 
and broad (30 - 50 km s$^{-1}$) methanol emission from the molecular 
ring. Two of the strongest methanol lines at 44 GHz are confirmed as 
masers by interferometric observations. These class I methanol masers 
are collisionally excited and are signatures of early phases of massive 
star formation in the disk of the Galaxy, suggesting that star formation 
in the molecular ring is in its early phase.  Close inspection of the 
kinematics of the associated molecular clumps in the HCN (J=1-0) line 
reveals broad red-shifted wings indicative of disturbance by 
protostellar outflows from young (few $\times 10^4\u yr $), massive 
stars embedded in the clumps. The thermal methanol profile has a similar 
shape, with a narrow maser line superimposed on a broad, red-shifted 
wing. Additional evidence for the presence of young massive protostars 
is provided by shocked molecular hydrogen and a number of striking 
ionized and molecular linear filaments in the vicinity of methanol 
sources suggestive of 0.5-pc scale protostellar jets.  Given that the 
circumnuclear molecular ring is kinematically unsettled and thus is 
likely be the result of a recent capture, the presence of both methanol 
emission and broad, red-shifted HCN emission suggests that star 
formation in the circumnuclear ring is in its infancy.

\end{abstract}

\keywords{Galaxy: center  - ISM: clouds- ISM - 
molecules - ISM: jets and outflows  - stars: formation}
\section{Introduction}
\label{introduction} 

The  circumnuclear molecular ring (CMR)  extends from 2 to 7 pc
 from the massive black 
hole at the center of the Galaxy (G\"usten et al. 1987; Jackson et al. 1993; 
Marshall, Lasenby \& Harris 1995; Latvakoski et 
al. 1999; Yusef-Zadeh et 
al. 2006; Bradford et al. 2005; Herrnstein  \&  Ho 2005; 
Sch\"odel et al. 2002; Ghez et al. 2005; Reid et 
al. 2004; Christopher et al. 2005, hereafter CSSY). 
The ring  was initially discovered more than 25 years ago (Becklin et al. 
1982; Gatley et al. 1986; Telesco et al. 2006) via detection of 
thermal emission from dust.  
Subsequent studies of molecular line emission 
showed  the ring to be inhomogeneous, clumpy, warped and kinematically 
disturbed (e.g., Jackson et al. 1993) suggesting that it has not fully settled  since 
the gas was captured.  
The existence  of the ring as well as the kinematics of molecular clouds in 
the Galactic center suggest that gas sporadically falls into the Galactic center and 
forms stars (Morris \& Serabyn 1996 and the references therein). Star formation in 
the ring itself has been 
thought to be 
inhibited by the strong tidal force in the gravitational potential of the black hole 
and nearby stars unless the gas density 
exceeds  $\sim 10^7 \ut cm -3 $. 
Recent high-resolution HCN (1-0) 
observations (CSSY)   infer that 
the clumps near the 
inner edge of the ring have densities of a few times higher than this limit, thus 
implying that star formation may  not be inhibited in the ring.  

Here we report the detection of H$_2$O and class I methanol masers in the 
CMR. 
Methanol masers are unambiguous signposts of on-going massive star 
formation throughout the Galaxy (Menten et al. 2002). Class I methanol masers are 
collisionally 
pumped, are relatively rare, and are often  offset from the associated  
young stellar objects (Sobolev et al. 2005; Kurtz et al. 2004; Voronkov et al. 
2004; Cesaroni 2005; Arce et al. 2007).  Interstellar H$_2$O masers are also 
collisionally 
excited (Elitzur,  Hollenbach \&  McKee 1992) at high densities of 
10$^{7-9}$ cm$^{-3}$ and are also found in star 
forming regions.  
Close inspection of HCN line profiles in the vicinity of the masers are suggestive of 
protostellar outflows. 
Together, these data provide evidence for
star formation activity in the CMR.

\section{Observations}

Using the Green Bank Telescope (GBT) of the National Radio Astronomy
Observatory\footnote{ The National Radio Astronomy Observatory is a facility
of the National Science Foundation operated under cooperative agreement by
Associated Universities, Inc.}, we searched for $7_0-6_1 A^+$ methanol I
(44.0694 GHz) and 6$_{16} - 5_{23}$ water masers (22.23508 GHz) toward the
CMR and the Sgr A East HII region near the center of the Galaxy.  Our
initial observations were made on 2007 October 30 using total power position
switching.  We used an uneven sampling covering 18 positions to map the
CMR in the water line and searched for methanol toward 10 selected molecular
clumps.  The spectrometer was configured with a 50 MHz bandwidth and 12.2 kHz
channels for the water observations and 200 MHz bandwidth and 24.4 kHz channels
for the methanol.  The channel spacing corresponds to 0.16 \kms\ in each case.
We removed residual baseline shapes by subtracting a polynomial fit to the
line-free channels.  Our typical noise after Hanning smoothing was 17 mJy per channel 
for the water
survey and 23 mJy per channel for the methanol survey.
We estimate the flux scale uncertainty to be about 15\%.
On 2008 March 25 we obtained deeper GBT observations of the methanol lines
detected in the first run.   We observed using frequency switching this time
to avoid contamination by sources in the reference beam.   We used a 200 MHz
bandwidth and 12.2 kHz channels.  The methanol spectra presented in this
paper come from the second set of observations. Also, 
a short  Very Large Array (VLA) observation of the 44 GHz methanol line 
with typical integration of 2-3 minutes was also carried 
out in the C configuration on 2008 May 25 using the 1A mode of the correlator with a 
velocity coverage 
of 84 \kms\ (12.5 MHz) and channel separation of 1.2 \kms\ (195 kHz). The phase calibrator 17443-31165 
was used every few minutes in fast switching mode to calibrate the three methanol sources detected in 
the CMR with the GBT.

\begin{figure}
\centering
\includegraphics[scale=0.6,angle=0]{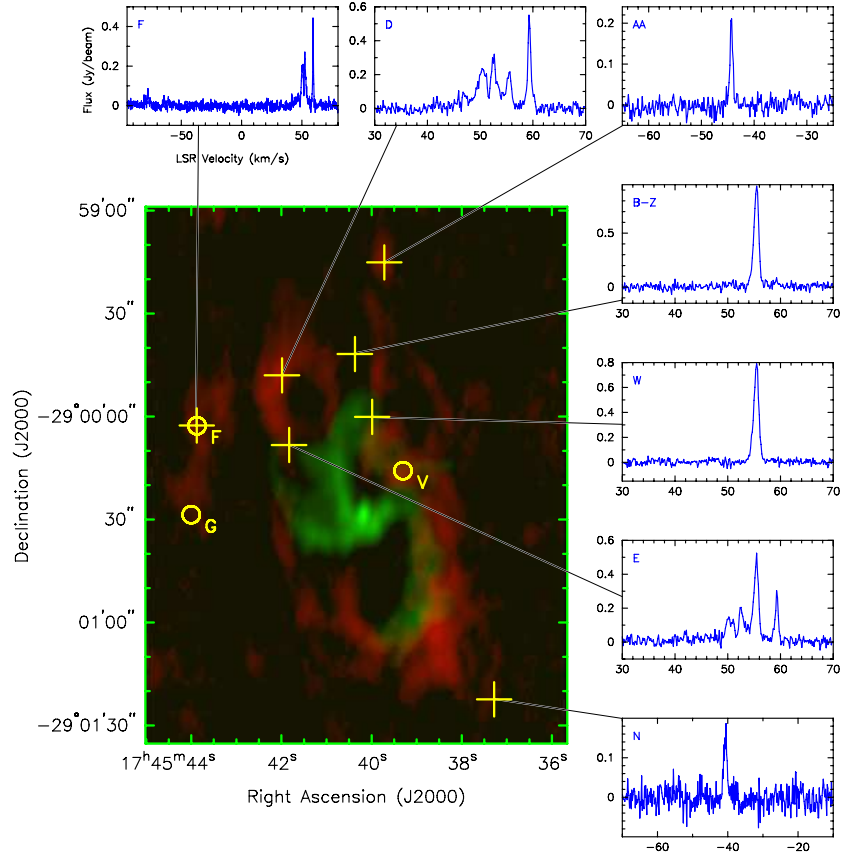}
\caption{
Integrated HCN line emission from CSSY (resolution 5.0$''\times2.7''$ at PA -2.7$^0$) is
shown in red. A  3.6 cm radio continuum image from the VLA (resolution 
4.7$''\times2.5''$
at PA -6$^0$)  is shown in green (Yusef-Zadeh \& Wardle 1993). 
 Crosses show the 
locations of
detected H$_2$O  and circles show locations of detected methanol, based on our GBT
observations.  The sizes of the symbols are three times smaller (for
clarity) than the beam used to observe them.  The panels surrounding the
image show H$_2$O spectra as observed with the GBT.  Labels on the spectra and
the figure match the clump designations of  CSSY.}
\end{figure}  

\section{Results}

Our search detected five methanol sources, three of which arise 
from the CMR and two from sources SgrA-A and SgrA-D of the SgrA East HII regions. 
Figure 1 shows the positions of detected methanol and water line emission from the 
CMR; the water spectra are shown in panels around the HCN image of the molecular 
ring.  A total of seven water line emission sources are detected toward the 
CMR. The three methanol detections coincide with HCN clumps F, G and V whereas water 
emission is detected toward clumps O, W, B-Z, D, E, F and an isolated clump which we 
label AA located at $\alpha, \delta \ (J2000); 17^h 45^m 39.4^s, -29^{\circ} 00' 19.5''$.
 A prominent `outer filament' to the north of this source is traced by both H$_2$ and 
HCN   
emission (Yusef-Zadeh et al. 2001; Bradford et al. 2005; CSSY).  
Interstellar water lines  are potentially contaminated by water lines 
from evolved OH/IR stars. Examination of the OH-IR survey by 
(Sjouwerman  \& van Langevelde 1996)  suggests this may be the case 
for the emission towards clump F, as 
stellar water maser
G359.956-0.05 
 is located at the 
half-power position of the GBT beam for the
positions
covering the northern and southern components of the clump (Fig. 1). A more detailed 
examination of interstellar water lines toward the CMR and methanol lines toward
Sgr A East HII regions will be 
given elsewhere. Here, we focus 
on the
three methanol maser sources that are detected toward clumps E, G and V in the CMR.

One of the methanol masers coincides with clump V  identified in the HCN 
emission from the molecular ring (CSSY).  Previous H$_2$ 
observations (Gatley et al. 1986) have 
indicated the presence of three shocked molecular H$_2$ features in the CMR, one of 
which coincides with clump V.  This clump is estimated  to have a hydrogen 
number density of 5.8$\times10^7$ cm$^{-3}$, a size of 0.1pc,  and a total mass of 
1.4$\times10^4$ \msol\ (CSSY). Figure 2 shows the methanol and HCN spectra toward 
two positions taken from the circled region where methanol line emission has been 
detected; these spectra are shown in  panels against the integrated HCN distribution 
of clump V.  The methanol spectrum shows a narrow line at 44.17  
\kms\ with a peak flux density of 0.682 Jy superimposed on broad emission that extends 
over LSR velocities between 40 and 100 \kms. 
The relatively large beam of GBT at 44 GHz (15$''$) means that we are  unable to 
distinguish 
whether the broad methanol emission arises from a cluster of weak, collisionally 
excited methanol masers or from thermal cores in star forming regions. Both types of 
emission have been identified in hot molecular cores with outflow activity in massive 
star forming regions such as Sgr B2 (Mehringer  \& Menten 1997)  and 
G31.41+0.31 (Kurtz et al. 2004). The narrow line with a FWHM$\sim0.35$ \kms\ 
strongly suggests the presence of class I methanol masers. 
Our VLA observations had insufficient sensitivity (56 mJy per channel) and spectral resolution (1.2 
\kms) to confirm the location within the GBT beam.

\begin{figure}
\centering
  \includegraphics[scale=0.6,angle=0]{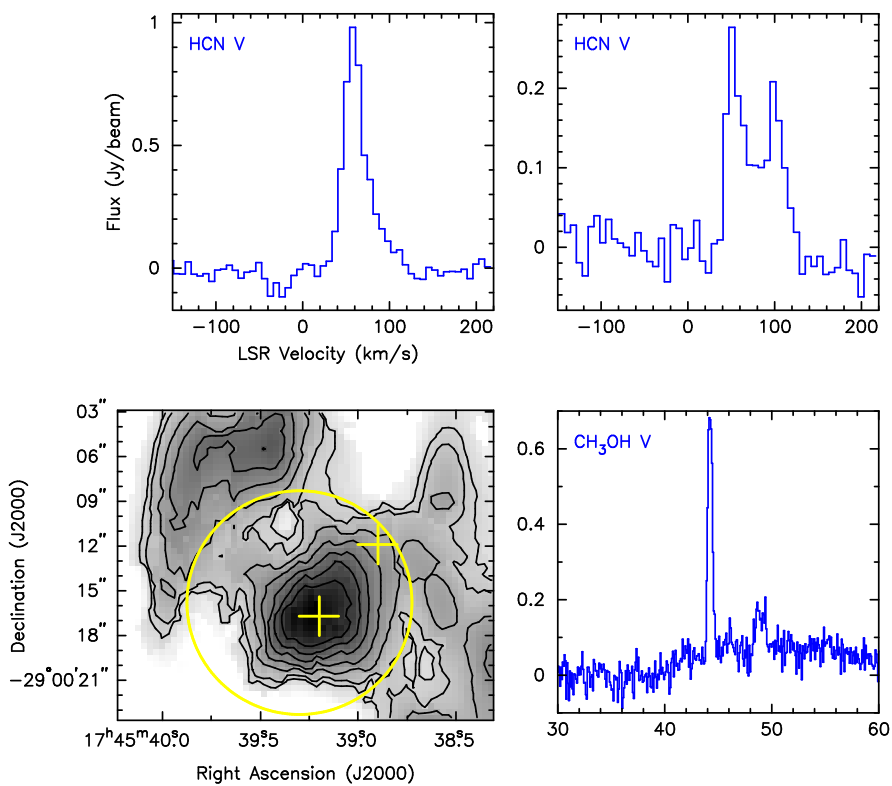}
\caption{{} 
(\textit{Bottom}) 
Contours of HCN emission from clump V are  shown to the left and  
the methanol spectrum at the position indicated by the 15$''$ circle (corresponding  to 
the  GBT beam at 44 GHz)  is shown to the right.
(\textit{Top}) 
 The left and right
panels  show
the HCN (1-0) emission profiles of the peak and northwestern
positions in clump V, respectively,  as indicated by crosses on the contour map. 
}
\end{figure}

The HCN spectra shown in Figure 2 consist of a peak (0.22 Jy) at 44.3 
\kms\ and a red-shifted component  at 100 \kms\ a few arcseconds to the 
northeast of the peak emission. The HCN spectra of the emission from the region 
corresponding to the 15$''$ GBT beam size of the detected methanol source shows 
clearly the presence of a red-shifted broad wing in its velocity profiles. The similarity 
of HCN and methanol line profiles toward clump V is a classic signature of one-sided 
molecular outflows in star forming regions (Plambeck \& Menten
1990).  The broad methanol emission is not widespread in the CMR as it 
appears to be located  within GBT beams where narrow lines are detected. 
The 
presence of shocked molecular 
gas traced by the H$_2$ 1--0 S(1) line at the position of clump V and collisionally 
excited methanol emission support this picture.  
At a deprojected distance of 1.1 pc from Sgr A* (CSSY), this clump requires  H$_2$ 
number density of 
2$\times10^7$ cm$^{-3}$ to be gravitationally bound, consistent 
with  density estimates based on excitation (CSSY) and within 
the range required to collisionally pump the methanol 
emission (Sobolev et al. 2005). The flux of methanol emission with its broad velocity  
$\sim30-50$ \kms\ are used to make an estimate of the column density of molecular 
gas in the 
quasi-thermal core. Using equation 6 of Mehringer \&  Menten (1997), we estimate 
the column density of 5.1$\times10^{15}$ cm$^{-2}$ using a typical 
flux 
of 200 mJy and a velocity width of FWHM$\sim25$ \kms\ and adopting a rotational temperature 
of 200K.    These thermal cores are only detected  where nonthermal
methanol maser components are detected. 
If we use a value of 10$^{-7}$ as the abundance ratio of CH$_3$OH to H$_2$, a column density
N$_{H_2}\sim5.1\times10^{22}$ cm$^{-2}$ is estimated. This corresponds to $\sim180$  
\msol\ of warm molecular gas associated with methanol masers. 
This low abundance of methanol is   consistent with recent 
observations showing  a lack of 96 GHz  methanol emission from  the 
CMR (Stankovi\'c et al. 2007) relative to the 20 and 50 \kms\ 
molecular clouds  adjacent to  the CMR.   These measurements suggest that
our detection of thermal methanol emission determined from the GBT spectrum is due to  
enhancement by 
localized  star formation 
rather than the intrinsic  physical conditions of  
the CMR, in contrast with  the 20 and 50 \kms\  molecular clouds.

The second methanol maser coincides with HCN clump F.  The
spectrum (Fig. 3)  shows three  narrow lines between 52 and 55 \kms\ with peak flux density 
of 2.38 Jy  at 50.2  \kms\
and a
broad red-shifted  wing extending up to 100 \kms, similar to the
velocity profile seen toward clump V. This is the only
one among the five detected methanol sources that shows a counterpart in
the 22 GHz water line. However,  as noted earlier, the OH-IR star
G359.956-0.05 (Sjouwerman  \& van Langevelde 1996) at the edge of the GBT beam 
 makes it difficult to identify any  interstellar
H$_2$O maser component associated with  clump F.  
The HCN emission has a   red-shifted wing near the  maser position (see Fig.  3).  
Our VLA observation of this source detected several weak components 
at 50 \kms\ with   flux densities  ranging between  
600  and 800 mJy  corresponding to antenna temperatures of 450 - 700 K and signal-to-noise of 5 and 7, 
respectively. 
The  VLA map with a resolution of 1.27$''\times0.56''$ 
(PA=14$^0$)  shows these weak sources to be 
distributed around a 20$''$ diameter  shell-like structure 
centered on   clump F. 
The narrow velocity component of the emission (line width 0.3 \kms), as measured by the 
GBT, increases  the brightness temperature of the weak VLA sources to 
1800 - 2800 K.


\begin{figure}
\centering
  \includegraphics[scale=0.6,angle=0]{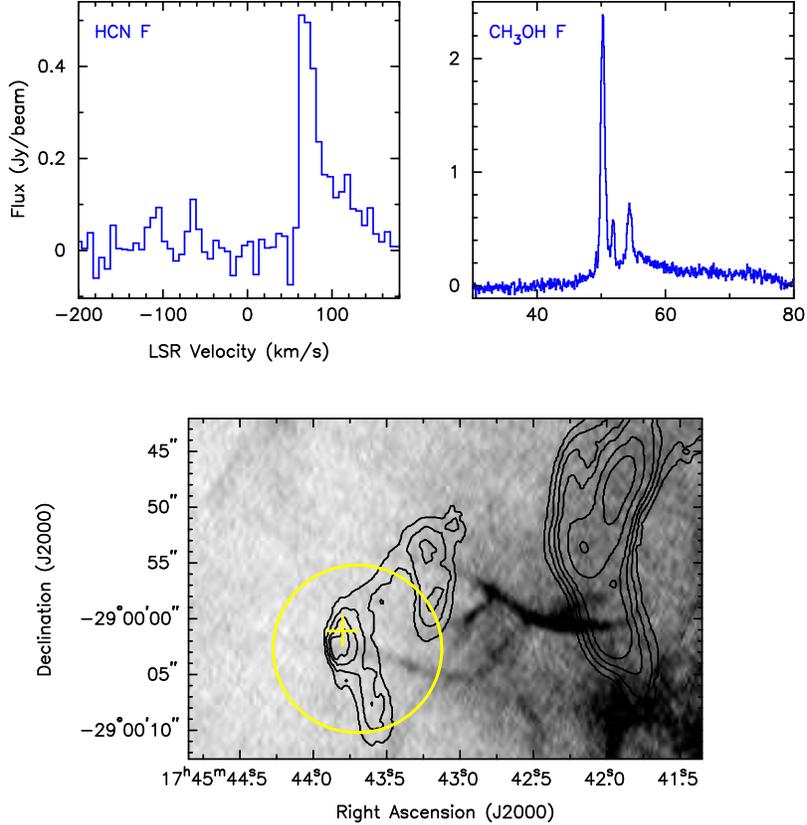}
\caption{
(\textit{Bottom}) 
Contours of HCN (1-0) emission are superimposed on a grayscale
continuum image at 6cm having  a resolution of 0.8$''\times0.5''$ (PA=11$^0$). 
Five linear features are evident  in
this  6cm continuum image, most of which appear to terminate at the
location of either a water or methanol source.  
The size of the circle 
corresponds to  the 15$''$ resolution  of the GBT beam at 44 GHz.
(\textit{Top}) 
The panels show the HCN and methanol emission  profiles 
from the peak position in clump F, as indicated  by a cross. 
}
\end{figure}  
  
Finally, the brightest methanol maser has  a peak flux 
density
3.03 Jy at 50.67 \kms\  and  is
associated   with clump G (Fig. 4). This again exhibits  a broad,  red-shifted line
profile, similar to those of clumps F and V. This profile 
peaks  near 50 \kms\ and extends up to 70
\kms.
VLA observations  detected a component  with a peak flux density of 
8 Jy  but was localized 
a few arcseconds outside the GBT 
beam at $\alpha, \delta (J2000) = 17^h 45^m 43^s.94, -29^{\circ} 00' 19''.4$. 
The higher VLA peak flux density than that of the  GBT 
suggests that the methanol 
source  is located beyond the half power of the GBT beam. 
The HCN spectrum showing a red-shifted wing in its velocity profile 
is detected 1.5$''$ northwest  of the location of the VLA spectrum, both of which are 
presented  in Figure 4. 
The brightness temperature of the VLA source using the 
0.3 \kms\ width  of the line is $\sim2\times10^4$K. 
The narrow velocity width and the high brightness temperature confirms 
the maser nature of the methanol emission detected  with the GBT.

Although methanol emission has been detected from three regions in the   
CMR, we find additional broad, red-shifted  HCN profiles, as in the region
in the southern lobe of the CMR at
$\alpha, \delta \; 17^h 45^m 38.45^s, -29^{\circ} 01' 0.28''$.
The line profile taken from a 0.4$''$ pixel  at this position peaks at -105 \kms\  
and
ranges  between -139 and -17 \kms.
This suggests that  the southern lobe of the CMR may  also be a site of
on-going star formation. Future mapping of the CMR in the methanol line 
will  be able to determine  if there are any methanol counterparts to 
red-shifted HCN line emission from the southern lobe.

\section{Discussion and Conclusion}

The presence of Class I methanol masers or thermal methanol
emission and the absence of ultra
compact HII regions in the immediate vicinity of clumps F and G are   
consistent with the earliest stages of massive star formation, within
a few $10^4 \u yr $ of the onset of gravitational collapse of a
molecular cloud core, and  prior to ionization by the UV flux from the
newborn massive star.  Furthermore, the HCN line profiles indicate
that  $\sim 100$\,\msol\ of molecular gas
has been swept up at $\sim50$  \kms\ presumably   by a  fast jet from  an 
embedded protostar.  This mass  is 
consistent with the estimate of molecular gas estimated from quasi-thermal component
of methanol emission.  
The momentum in the swept up gas  is within
the range associated with massive protostars in star forming regions (Arce et al. 2007)
in the disk of the Galaxy.  Assuming a jet speed of 500 \kms\ and an   
age of $10^4$ years, the inferred mass-loss rate from the protostar is  
$1\ee -3 $ \msol yr$^{-1}$, again consistent with the range observed for
massive protostars in the Galactic disk (Arce et al. 2007).

The presence of components with negative and positive velocity relative to the bulk of clump 
G indicates a bipolar outflow.  The one-sided, redshifted profiles of clumps F and V suggest 
that in these cases the approaching side of the jet has burst through the edge of the 
molecular clump.  Jets emerging from cloud cores in the Galactic center would become exposed 
to the UV radiation of hot stars, thus the full extent of the jet could in principle be 
visible as an ionized source.  Indeed, radio  continuum image of the inner few arc-minutes 
shows several ionized linear features associated with molecular clumps in the CMR 
or the maser sources we have detected.  A more detailed discussion of these ionized 
jet-like features will be given elsewhere.

\begin{figure}
\centering
  \includegraphics[scale=0.5,angle=0]{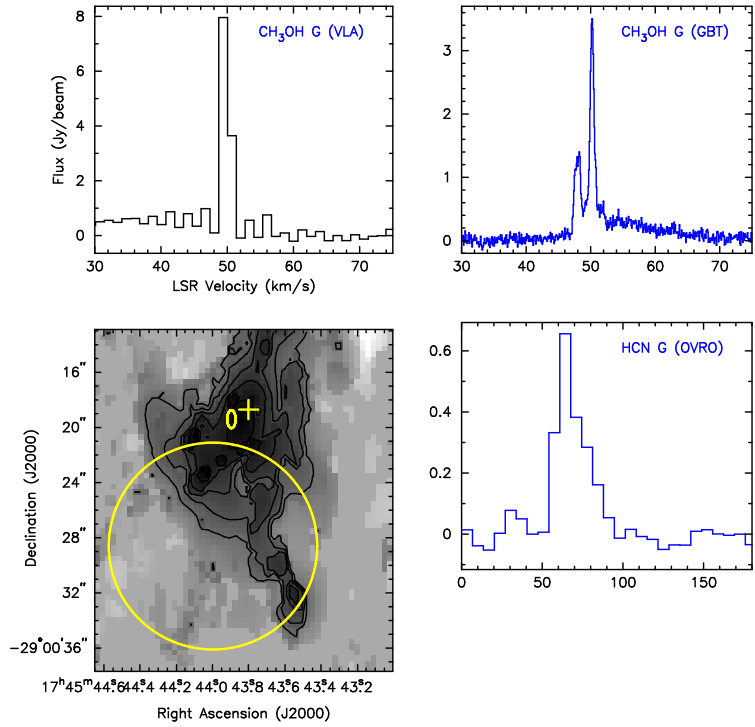}
\caption{
(\textit{Bottom Left}) 
Contours of HCN emission 
from  clump G (CSSY). The size of the circle corresponds to the 15$''$ resolution 
of the GBT beam at 44 GHz. 
Using OVRO, VLA and GBT,  the panels   show the HCN and CH$_3$OH profiles of emission from a pixel size 
of 
0.4$''$ for OVRO, ($1.3''\times0.56''$,  PA=7$^0$) for VLA   and  
15$''$ for GBT,  as indicated by the sizes of the cross, the ellipse and the  circle, respectively. 
}
\end{figure}

Our detection of collisionally excited methanol masers, the broad molecular lines and 
shocked H$_2$ line associated with the clumps of gas indicates the presence of outflows (Plambeck \& 
Menten 1990) from 
massive protostars in the CMR 
with ages of a few $10^4$  yr.  The lack of ultracompact HII regions throughout 
the ring implies that older, massive protostars are not present in appreciable 
numbers, and therefore that star formation in the CMR is only just beginning. This is 
consistent with the youthful dynamical state of the ring; the high velocity 
dispersion (Jackson et al. 1993)  ($\sim$30 \kms) in the ring suggests that it is 
marginally 
gravitationally unstable. The velocity dispersion is expected to be damped by 
clump-clump collisions on a time scale 
comparable to the orbital 
time of the gas $\sim 10^5 \u yr $.  It is also consistent with the high densities recently inferred 
from HCN measurements (CSSY), which imply the existence of 
gravitationally bound 
clumps, the precursors of collapse. Furthermore, the asymmetry in the northern and 
southern halves of the CMR suggests that the southern half 
of the CMR has not begun its star formation activity.

The youth of the ring supports the idea that star formation within the central few 
parsecs of the Galaxy is continually being fed by the capture of interstellar gas 
clouds.  This process may be responsible for the unusual stellar populations in the 
central few pc of the Galaxy, which include very young stars (Paumard et al. 
2006; Lu et al. 2006) (ages $\sim$  
few Myr), as well as a population built up by continuous formation of stars over 
the last billion years.  We have recently proposed (Wardle \& Yusef-Zadeh 2008) 
that 
the sub-parsec-scale 
disk of massive stars orbiting the massive black hole at the Galactic center and the 
CMR can be created by partial accretion of extended Galactic center clouds when they 
envelop Sgr A* on a passage through the inner Galactic center.  The cancellation of 
angular momentum of the captured cloud by self-interaction naturally creates a 
compact, gaseous disk of material close to Sgr A* in which star formation 
subsequently takes place.  The CMR may be a relic from a recent passage of a cloud 
similar to that envisioned for the young stellar disk near Sgr A*, but with a lower 
incident cloud resulting in the capture of a larger region of the incoming cloud. 
Future studies of the molecular and ionized material and protostars associated with 
the CMR should be able to examine the role that this young cloud makes in our 
understanding of the evolution of gas and stars in in the nucleus of our Galaxy.

Acknowledgments:  We are extremely  grateful to M. Christopher and N. Scoville
for providing us with their HCN data. We also thank the referee for careful reading
of this paper.

\newcommand\refitem{\bibitem[]{}}


\begin{thebibliography}{99}

\bibitem{20}
Arce, H. G. et al.  2007, 
in Protostars and Planets V, B. Reipurth, D. Jewitt, and K. Keil (eds.), University
of Arizona Press, Tucson, 245


\bibitem{11} Becklin, E., Gatley, I.  \& Werner, M. W. 1982, 
{\it ApJ},  {\bf 258}, 135

\bibitem{8}
        Bradford, C. M., Stacey, G. J., Nikola, T., Bolatto, A. D.,  Jackson, J. M.,
Savage, M. L.,  Davidson, J. A. 2005, 
{\it ApJ},  {\bf623}, 866

\bibitem{19} Cesaroni, R. 2005, 
Outflow, Infall, and Rotation in High-Mass Star Forming Regions,
{\it Ap.Space.Science },  {\bf 295 }, 5

\bibitem{10} Christopher, M. H.,  Scoville, N. Z.,  Stolovy, S. R. \& Yun, M. S. 
2005, 
{\it ApJ},  {\bf 622}, 346 (CSSY)

\bibitem{22} Elitzur, M.,  Hollenbach, D. J. \&  McKee, C. F. 1992, 
{\it ApJ},  {\bf 394}, 221




\bibitem{12}  Gatley, I., Beattie, D. H., Lee, T. J., Jones, T. J.
         \& Hyland, A. R. 1986, 
{\it MNRAS},  {\bf 222},  299

\bibitem{2}
Ghez, A. M., et al. 2005,  {\it ApJ},  {\bf 620}, 744


\bibitem{4}
G\"usten, R., Genzel, R., Wright, M. C. H., Jaffe, D. T. 1987, 
{\it ApJ},  {\bf 318}, 124

\bibitem{9}  Herrnstein, R. M. \&  Ho, P. T. P. 2005
{\it ApJ},  {\bf 620}, 287

\bibitem{5}  Jackson, J. M.,
Geis, N., Genzel, R., Harris, A. I., Madden, S.,
Poglitsch, A., Stacey, G. J. \& Townes, C. H.  1993, 
{\it ApJ},  {\bf 402}, 173


\bibitem{17} Kurtz, S. \& Hofner, P. \& Ãlvarez C. V. 2004, 
{\it ApJS},  {\bf 155}, 149

\bibitem{6}
Marshall, J., Lasenby, A. N. \& Harris, A. I. 1995,
{\it MNRAS}, {\bf 277}, 594

\bibitem{6} Latvakoski, H.M., Stacey, G.J., Gull, G.E. \& Hayward, T.L. 1999, 
{\it ApJ},  {\bf 511}, 761

\bibitem{30} Lu, J. R., Ghez, A. M., Hornstein, S. D., Morris, M., Matthews, K. \&
Thompson, D. J. 2006, 
{\it JPCS}, {\bf 54},  279

\bibitem{24}
Mehringer, D. M. \& Menten, K.  M. 1997, 
{\it ApJ},  {\bf 474 }, 364


\bibitem{15}
Menten, K. M. et al. 
1992,  {\it ApJ},  {\bf 401}, L39



\bibitem{14}
Morris, M. \& Serabyn, E.  1996, {\it
ARAA}, {\bf 34},  645


\bibitem{29}
Paumard, T. et al. 2006, 
{\it ApJ},  {\bf 643}, 1011


\bibitem{25}    Plambeck, R. L. \& Menten, K. M. 1990, 
{\it ApJ},  {\bf 364}, 555 

\bibitem{3}  Reid, M. J.  \& Brunthaler, A. 2004, 
{\it ApJ},  {\bf 616}, 872


\bibitem{1}  Sch\"odel, R., Ott, T., Genzel, R., Eckart, A., Mouawad, N., \& 
Alexander,
T. 2002, 
{\it Nature},  {\bf 419}, 694



\bibitem{23} 
Sjouwerman, L. O. \& van Langevelde, H. J. 1996, 
{\it ApJ},  {\bf 461}, L41

\bibitem{23} Sobolev, A. M. et al. 2005,  
in the proceedings of the IAUS 227, eds: Cesaroni, R. Churchwell, E. B.,
Felli, M., Walmsley, C. M. 174



\bibitem{16}
Stankovi\'c, M., Seaquist, E.~R.\,   M{\"u}hle, S.  Leurini, S. \&  Menten, K.~M. 
2007, 
in ``Molecules in Space and Laboratory'', , May 14-18, 
2007.  Eds: J.L. Lemaire, et al.  p.15

\bibitem{13}  Telesco, C. M., Davidson, J. A. \& Werner, M. W. 1996, 
{\it ApJ},  {\bf 456}, 541



\bibitem{18} Voronkov, M. A.,  Brooks, K. J.,  Sobolev, A. M., Ellingsen, S. P.,  
Ostrovskii,
A.  B. \& Caswell, J. L. 2006, 
{\it MNRAS},  {\bf 373 }, 411

\bibitem{27}
Wardle, M. \& Yusef-Zadeh, F. 2008, 
{\it ApJ}, submitted (arXiv:0805.3274)


\bibitem{21} Yusef-Zadeh, F.,  Hewitt, J. W. \& Cotton, W. 2004, 
{\it ApJS},  {\bf 155}, 421


\bibitem{7}  Yusef-Zadeh, F.,  Stolovy, S. R.,  Burton, M., Wardle, M. \& Ashley, M. C. B. 
2001,  {\it ApJ},  {\bf 560}, 749




\end{thebibliography}
\end{document}